\documentclass{aa}
\usepackage{txfonts}
\usepackage{graphicx}
\usepackage{amsmath}
\usepackage{natbib}
\usepackage{amssymb}
\bibpunct{(}{)}{;}{a}{}{,} 
\begin{document}
\title{Scale invariant cosmology II:  model equations and properties }
\titlerunning{Scale invariant cosmological models}

\author{Andr\'e Maeder
}
\authorrunning{Maeder}

\institute{Geneva Observatory, Geneva University, CH--1290 Sauverny, Switzerland\\
email: andre.maeder@unige.ch
}

\date{Received  / Accepted }


\abstract
{}
{We want to establish the basic properties  of a scale invariant   cosmology, that
also accounts  for the hypothesis of  scale invariance of the empty space at large scales. 
}
{We write the basic  analytical properties of the scale invariant cosmological models. }  
{The hypothesis of  scale invariance  of the empty space at large scale brings interesting simplifications in the scale invariant
 equations for cosmology. There is one new term, depending 
on the scale factor of the scale invariant cosmology,  that  opposes to gravity and favors an accelerated expansion.
We first consider a  zero-density model and find an accelerated expansion, going like $R(t) \, \sim \, t^2$. 
In models with matter present,
the displacements due to the new term  make a significant contribution
 $\Omega_{\lambda} $ to the energy-density of the Universe, satisfying an equation 
of the form $ \; \Omega_{\mathrm{m}}+\Omega_{\mathrm{k}}+\Omega_\lambda =1 $.

 Unlike the Friedman's models, there is a whole family of flat models ($k=0$) 
with different density parameters $\Omega_{\mathrm{m}} < 1$.
We examine the basic relations between the  density and geometrical properties, as well as the conservation laws.
 The models containing matter have an inflexion point, with first a braking phase followed
by an accelerated expansion phase.
}
{The scale invariant models have  interesting properties and deserve further investigations  }


\maketitle
 
 \section{Introduction}

The questions regarding the acceleration of the Universe expansion and the dark energy
 dominate the cosmological research for about two
decades  \citep{Wein89,Carr92,Riess98,Perl99,Frie08,Feng10,Port11,Sola13}.
The situation is like if an interaction of unknown nature opposes the gravitation at cosmological scales. 
 A high number of different hypotheses have been formulated to try to explain the accelerated expansion.

 There has been long-standing efforts to build a theory of gravitation, which also include scale invariance in addition to the invariance to a general coordinate transformation \citep{Weyl23,Eddi23,Dirac73,Canu77}. 
 Scale  covariant  theories were often developed  for trying
to support the view that the gravitational constant $G$ varies with time
in relation with the so-called Large Number Hypothesis  \citep{Dirac73}. Here, we do not follow this hypothesis and  
the gravitational constant remains a constant.

The laws of physics generally are not scale invariant, since the matter content of the medium considered 
may fix some scales of mass, length and time. However, the empty space as it is considered for example in the Minkowski
metric  has no preferred scale and in Paper I we have made the assumption that the empty space is scale invariant at large scales.
 This has  lead to two differential relations
between the cosmological constant and the scale factor $\lambda$, which expresses how the line element may change with time.

It is well known that at the quantum level, the properties of the vacuum are not scale invariant,
 since quantum physics defines units of length,
 time and mass. However, at the level of the Universe, especially in view of the problem of the accelerated expansion and dark
energy, we really do not know whether the assumption of scale invariance applies or not. This is what we want to explore 
on the basis of  general  field equations containing   scale  invariance in addition  to the usual invariance 
to the group of transformations of  curvilinear coordinates in Riemann space, which characterizes General Relativity.   This enlarges
the group of invariances sub-tending  the theory of gravitation, with an invariance also present in electromagnetism.
General Relativity appears as particular case of the new approach, when the scale factor is kept constant through space-time. 

 We  make an explicit use of  the  two differential relations derived from the assumption of scale invariance of the empty space
at large scales. These relations play an essential role and  lead to  solutions showing cosmic acceleration.

Sect. 2 gives the basic equations of  scale invariant cosmology.
In Sect. 3, we examine the case of an empty Universe in the scale invariant context.
The critical density, the $\Omega$ and geometrical parameters are studied in Sect. 4.
 In Sect. 5, the appropriate  conservation laws for a scale invariant cosmology are derived.
Sect. 6 contains the conclusions.

\section{The  equations of scale invariant cosmology and  properties}

The scale or gauge invariant cosmology assumes that the field equations are invariant to a transformation of the line element like
$ds' \, = \, \lambda(x^{\mu}) \, ds $,
where $ds'\,^2  =  g'_{\mu \nu} dx^{\mu} \, dx^{\nu} $ is the line element in General Relativity, while  $ds^2  =  g_{\mu \nu} dx^{\mu} \, dx^{\nu} $  
is the line element in a  more general framework where scale invariance is supposed to also  be a fundamental property.
The quantities in the framework of General Relativity are noted with a prime, while those in the 
more general framework, that includes scale invariance, are  without a prime.
  The parameter  $\lambda$ is the scale factor connecting the two line elements. According to the Cosmological Principe of homogeneity and isotropy, $\lambda$  can only depend on the cosmic  time $t$.
 
The scale invariant field equation in cotensorial form  has been given  in Eq. (21)  of Paper I.
The general field  equation has been  applied to the empty space  with the Minkowski metric.
This has lead to two differential equations, which will play an essential role in the present work,

\begin{equation}
\  3 \, \frac{ \dot{\lambda}^2}{\lambda^2} \, =\, \lambda^2 \,\Lambda_{\mathrm{E}}  \, , \quad  \quad \mathrm{and} \quad \quad
2 \, \frac{\ddot{\lambda}}{\lambda} - \frac{ \dot{\lambda}^2}{\lambda^2} \, =\, \lambda^2 \,\Lambda_{\mathrm{E}}   \, .
\label{diff1}
\end{equation}
\noindent
They can also be written in equivalent forms,

\begin{equation}
\frac{\ddot{\lambda}}{\lambda} \, = \,  2 \, \frac{ \dot{\lambda}^2}{\lambda^2} \quad \quad \mathrm{and} \quad \quad
 \frac{\ddot{\lambda}}{\lambda} -\frac{ \dot{\lambda}^2}{\lambda^2} \, = \, \frac{\lambda^2 \,\Lambda_{\mathrm{E}}}{3} \, .
\label{diff2}
\end{equation}

\noindent
The dots indicate the time derivatives and $\Lambda_{\mathrm{E}}$ is the Einstein cosmological constant, (we do not assign a prime to
it, since there is here no ambiguity).
These  relations, derived from the hypothesis of  the scale invariance of  the empty space at large scales, express some
interesting results: 
\begin{itemize}
\item There is a  relation of the  cosmological constant $\Lambda_{\mathrm{E}}$ with the scale factor $\lambda$ and its derivatives.
\item There may be an energy-density  associated to  the time-variations  of the scale factor.
\item The first of equations $(\ref{diff2})$   gives the  time dependence of $\lambda (t)$,
\begin{equation}
\lambda \, = \, \sqrt{\frac{3}{\Lambda_{\mathrm{E}}}} \, \frac{1}{c \,t}  \, .
\label{lamb}
\end{equation}
\end{itemize}
\noindent
If we choose $\lambda$ to be unity at the present  time  $t_0$, then we have $\lambda \, =  \, t_0 /   t \,$.
As pointed out in Paper I,  the first of equations ({\ref{diff2}) does not imply a particular origin for the time $t$. The origin will depend on the model considered. This also means that the amplitude of the variations of $\lambda(t)$ over the evolution of the Universe, from 
the origin to now, will strongly depend on the considered cosmological model.

The metric appropriate  to cosmological models is the Robertson-Walker metric, characteristic of the homogeneous and isotropic
space.  
A first step towards the equations we want to use can be derived in various equivalent ways  \citep{Canu77}: -- by  expressing
the  general cotensorial field equation with the Robertson-Walker metric,  -- by taking advantage that there is a conformal transformation
between the metrics $g'_{\mu \nu}$ and $g_{\mu \nu}$, --  by applying a scale transformation 
 to the current equations of cosmologies in $R(t), \,  \dot{R}, \, \ddot{R}$. 
The details of this third possibility  for obtaining  the basic equations are given in Appendix A. 
These equations are,

\begin{equation}
\frac{8 \, \pi G \varrho }{3} = \frac{k}{R^2}+
\frac{\dot{R}^2}{R^2}+ 2 \, \frac{\dot{\lambda} \, \dot{R}}{\lambda \, R}+\frac{\dot{\lambda}^2}{\lambda^2} - \frac {\Lambda_{\mathrm{E}} \lambda^2}{3} \,
\label{E1p}
\end{equation}

\noindent 
and

\begin{equation}
-8 \, \pi G p = \frac{k}{R^2}+ 2 \frac{\ddot{R}}{R} + 2 \frac{\ddot{\lambda}}{\lambda}+\frac{\dot{R}}{R}^2
+ 4 \frac{\dot{R} \, \dot{\lambda}}{R \, \lambda}-\frac{\dot{\lambda^2}}{\lambda^2} -\Lambda_{\mathrm{E}} \,  \lambda^2  \, .
\label{E2p}
\end{equation}

\noindent
These equations contain several additional terms with respect to the standard case.
In the same way as $\Lambda_{\mathrm{E}}$, which is related to the energy-density  of the vacuum, intervenes in the
$\Lambda$CDM model, expressions (\ref{diff1}) and (\ref{diff2}) for the empty space, which  characterize $\lambda$ 
 and its properties also apply. 
Thus, with   (\ref{diff1}) and (\ref{diff2}), the two above cosmological equations (\ref{E1p}) and
(\ref{E2p}) may be simplified and become,

\begin{equation}
\frac{8 \, \pi G \varrho }{3} = \frac{k}{R^2}+\frac{\dot{R}^2}{R^2}+ 2 \,\frac{\dot{R} \dot{\lambda}}{R \lambda} \,
\label{E1}
\end{equation} 

\noindent
and 
\begin{equation}
-8 \, \pi G p  = \frac{k}{R^2}+ 2 \frac{\ddot{R}}{R}+\frac{\dot{R^2}}{R^2}
+ 4 \frac{\dot{R} \dot{\lambda}}{R \lambda}  \, .
\label{E2}
\end{equation}
\noindent
The combination of these two equations  leads to

\begin{equation}
-\frac{4 \, \pi G}{3} \, (3p +\varrho)  =  \frac{\ddot{R}}{R} + \frac{\dot{R} \dot{\lambda}}{R \lambda}  \, .
\label{E3}
\end{equation}
\noindent
There, $G$ is the gravitational  constant, a real constant,  $k$ is the curvature parameter which  takes values $0$ or $\pm 1$,  $p$ and $\varrho$ are the pressure and density in the scale invariant system of coordinates. Einstein cosmological constant has disappeared from
the equations due to the account of the properties of the vacuum at macroscopic or large scales.
Interestingly enough, these three equations differ from the classical ones, in each case  only by the presence of  one additional 
terms containing  $ \dot{R} \, \dot{\lambda}/ (R \, \lambda)$. This additional term is different from zero if $\lambda(t)$ is not
a constant.  Indeed, if $\lambda(t)$ is a constant, one gets the usual equations of cosmologies for the expansion term $R(t)$. 
This means that at any fixed time, the effects which do not depend on the time evolution of the Universe, are just those 
predicted by General Relativity.

 What is the significance of  this additional term?
Let us consider (\ref{E3}). The  term on the left represents the attractive gravitational potential due  to the
matter and energy  present in the considered model. 
This term contributes negatively to the second derivative of $R$ and  thus  produces a braking of the motion
of the comoving particles. 
The second term on the right of (\ref{E3}) is negative, since  according to
(\ref{lamb}) we have 
\begin{equation}
\dot{\lambda}/\lambda \, = \, - \frac{1}{t} \, . 
\label{lambdot}
\end{equation}
\noindent
This second term 
  represents  {\emph{ an acceleration that opposes the gravitation}},  it depends on both the
Hubble constant $\dot{R}/R$ and on the relative change $\dot{\lambda}/\lambda$
of the scale factor. This term may have a significant effect on the evolution of the Universe producing  an 
acceleration, which may particularly reveal itself during the advanced stages of evolution,
since according to (\ref{E3}) the acceleration is proportional to the relative velocity of the expansion. 
Equations (\ref{E1}) to (\ref{E3})   incorporate the scale invariance of the field equations as well 
the scale invariance properties of the vacuum at large scales.
Numerical models have to be calculated to
provide the solutions corresponding to the various choices of  density and curvature parameters.

\section{The case of an empty Universe} \label{empty}

Let us first consider in the scale invariant framework the interesting  case of a  non-static model Universe with no matter 
nor significant radiation ($\varrho$ = 0  and $p$ = 0). The corresponding model in the standard case would be the
empty Friedman model, (such a model with $k=-1$ has an expansion given by $R(t) \sim t$).
 We note that empty models  are interesting as they represent the asymptotic limit
of models with lower and lower densities. Moreover, we may be  confident that empty models
should be scale invariant, since there is no matter present able to provide any scale.
Expression (\ref{E3}) becomes simply

\begin{equation}
 \frac{\ddot{R}}{R} = - \frac{\dot{R} \dot{\lambda}}{R \lambda}  \, .
\label{E3vide}
\end{equation}

\noindent
The integration with (\ref{lambdot}) gives $\dot{R}=a \,t$, where $a$ is a constant, a further integration gives

\begin{equation}
R = a(t^2-t^2_{\mathrm{in}}) \,  .
\label{rt2}
\end{equation}

\noindent
$R(t)$ grows like $t^2$, the initial instant $t_{\mathrm{in}}$ of the model  is  chosen at the origin  $R(t_{\mathrm{in}})=0$  of the 
considered model, ($t_{\mathrm{in}}$ is not necessarily 0). The model is non-static and 
the Hubble value  at time $t$ is 

\begin{equation}
H = \frac{\dot{R}}{R} = 2  \frac{t}{(t^2-t^2_{\mathrm{in}})} \,  
\label{H}
\end{equation}

\noindent
We do not know yet $t_{\mathrm{in}}$, however we may use  Eq. (\ref{E1}) and get,

\begin{equation}
\dot{R}^2 \, t - 2 \, \dot{R}  R + k t =0   \, ,
\end{equation}

\noindent
This equation   leads to a second  expression of the Hubble constant,

\begin{equation}
H=\frac{\dot{R}}{R}= \frac{1}{t} \pm \frac{\sqrt{1- k\frac{t^2}{R^2}}}{t} \, .
\label{root}
\end{equation}

\noindent
For an empty model, we may take $k=-1$  or $k=0$ (this is consistent with (\ref{kom})  in the study of the geometrical parameters below).
Let us first consider the case $k=-1$.
The dimensions of $k$ go like $[R^2/t^2]$. Fixing the scale so that  $t_0 =1$ and $R_0=1$ at the present time, we get from
(\ref{root}), the present Hubble constant $H_0$ being positive,

\begin{equation}
H_0= \frac{1+\sqrt{2}}{t_0} \,.     
\label {Hovide}
\end{equation}

\noindent
For $k=-1$,
the above value represents a lowest bound  of $H_0$-values,  expressed as a function of $t_0=1$,  to the  models with
non-zero densities, (since the steepness of  $R(t)$ increases with higher densities according to (\ref{E1})).
As is usual, the  value of $H_0$ should be expressed in term of the age  $\tau= t_0 -t_{\mathrm{in}}$ of the 
Universe in the considered model. 
We find $ t_{\mathrm{in}}$ by expressing the equality of the two values of $H_0$ obtained by (\ref{H}) and (\ref{Hovide}),

\begin{equation}
\frac{t_{\mathrm{in}}}{t_0} \,= \,  \sqrt{2} - 1 \, .
\end{equation}
\noindent 
This is  the minimum value of $t_{\mathrm{in}}$, (we notice that here the scale factor $\lambda (t)$ at the origin has a  value limited 
to $1+\sqrt(2)=2.4142$).
The corresponding age $\tau$ becomes $\tau = (2-\sqrt{2}) \, t_0$  and  we may now 
 express the  value of $H_0$ from (\ref{Hovide}) as a function of the  age $\tau$ of the Universe.
We have quite generally, indicating here in parenthesis the timescale referred to,

\begin{equation}
\frac{H_0(\tau)}{\tau} \, = \, \frac{H_0(t_0)}{t_0} \, , \quad  \mathrm{thus} \; H_0(\tau) \, =   \, H(t_0) \, \tau  \, .
\label{htau}
\end{equation}
\noindent
Thus, we get the following value of $H_0 (\tau)$, which is a maximum value, resulting from the fact 
that $\tau$ is a maximum,  
\begin{equation}
H_0 (\tau)\,  = \, \sqrt{2} \,  \quad \mathrm{for} \; \; k= -1 \, .
\label{hovide}
\end{equation}

Let us now turn to the  empty model with  $k=0$. We  have  according to (\ref{root})

\begin{equation}
H_0 \, = \, \frac{2}{t_0} \, ,
\label{hkmin}
\end{equation}
\noindent
 As for $k=-1$, this value for the empty space is the minimum value of $H_0$ expressed
in the scale with $t_0=1$.
The comparison  of (\ref{hkmin}) and ({\ref{H}) leads
to $t_{\mathrm{in}}=0$ and thus $ \tau=t_0$. Here also, $t_{\mathrm{in}}$ is the minimum value for all  models with $k=0$
and thus  $\tau$ is the longest  age.
We have, expressing $H_0$ as a function of $\tau$,

\begin{equation}
H_0(\tau) \, = \, 2 \, , \quad \mathrm{for} \; \; k= 0 \, .
\label{hovidezero}
\end{equation}
\noindent
Here also, $H_0(\tau)$ is an upper bound  for the models with $k=0$, due to the fact that the above  $\tau$ is a maximum.
We see that  the empty models, whether $k=-1$ or $k=0$, obey very simple properties.

The empty  scale invariant model  Universe  expands  like $t^2$,  thus with a strongly accelerating expansion over the ages. 
It expands much more rapidly than the corresponding Friedman model, which experiences   a linear expansion
$R \sim t$, with  $H_0 = 1/t $, and shows no acceleration. Thus,  here the effects of scale invariance appear as the source
of a strongly accelerated expansion, consistently with the remark made above about relation (\ref{E3}).

\section{Cosmological properties and parameters}

We now examine some general properties and interesting parameters of the cosmological models
based on equations (\ref{E1}) - (\ref{E3}).

\subsection{Critical density and $\Omega$-parameters}   \label{dens}

 The critical density  corresponding to the case $k=0$
of the flat space  is an essential model reference. Since the basic equations
are different from the Friedman models, the critical density is also defined by  a different
expression. From (\ref{E1}) and (\ref{lamb}), we have

\begin{equation}
\frac{8 \, \pi G \varrho^{*}_{\mathrm{c}} }{3} =H^2 -2\, \frac{H}{t} \, .
\label{rocm1}
\end{equation}

\noindent
 We mark with a  *  this critical density that does not
correspond to the usual definition,

\begin{equation}
\varrho^{*}_{\mathrm{c}} =  \frac{3 \, H^2}{8 \pi G}  \left(1- \frac{2}{t \, H }\right)  \, .
\label{roc}
\end{equation}

\noindent
Expression (\ref{roc}) evidently also applies for 
the critical density at present time $t_0$, with  a Hubble value $H_0$.
 This critical density (\ref{roc})  is smaller than  
the corresponding critical density of Friedman models with $k=0$.
The parenthesis in (\ref{roc}) is always positive. This is true at any time $t$, since $2/(t\, H)= 2 (dt/t) (R/dR)$
and the relative  growth rate  for non empty models  is  higher than $t^2$.
Indeed, we have seen that models satisfying relation (\ref{E1}) and with $k=0$  (resp. $k=-1$) 
 have a value of  $H_0 \geq \frac{2}{t_0}$}
(resp. $H_0 \geq \frac{1+\sqrt{2}}{t_0}$})  according to (\ref{hovidezero})  (resp. (\ref{hovide}). Thus,   we have, for $k=0$,
$\frac{2}{t_0 \, H_0} \, \leq  1$ and, for $k=-1$, 
$\frac{2}{t_0 \, H_0} \, \leq \ \frac{2}{1+\sqrt{2}} =0.828$, so that
 the parenthesis is  zero or positive.

 Let us now examine the various contributions to the  mass and energy.
Expressing (\ref{E1}) at  time $t$ and dividing by $H^2$, we get with (\ref{lamb})

\begin{equation}
\frac{8 \, \pi G \varrho}{3 H^2} - \frac{k}{R^2 H^2}+ \frac{2}{H t} \, = \, 1 \, .
\label{smm}
\end{equation}

\noindent
If we now introduce the expression (\ref{roc}) for $\varrho^*_{\mathrm{c }}$, we get

\begin{equation}
\frac{\varrho}{\varrho^*_{\mathrm{c }}} - \frac{k}{R^2 H^2} + \frac{2}{H t}
\left(1 - \frac{\varrho}{\varrho^*_{\mathrm{c }}} \right)= \, 1 \, .
\label{sum}
\end{equation}

\noindent
With the definitions,

\begin{equation}
\Omega^*_{\mathrm{m}} = \frac{\varrho}{\varrho^*_{\mathrm{c }}}  \, , \quad \mathrm{and} \quad
\Omega_{\mathrm{k}} = -\frac{k}{R^2  H^2} \;  , 
\label{defo}
\end{equation}

\noindent
expression (\ref{sum}) becomes

\begin{equation}
\Omega^*_{\mathrm{m}} \, + \, \Omega_{\mathrm{k}} \, + \frac{2}{H \, t} \left( 1 - \Omega^*_m \right) \, = \, 1  \, .
\label{Omega}
\end{equation}

\noindent
The quantity $\Omega^*_{\mathrm{m}}$ is the ratio of the density to the critical density in the framework of the scale invariant theory.
We see that   $\Omega^*_{\mathrm{m}}=1$    implies $\Omega_{\mathrm{k}}=0$ and reciprocally, consistently with the
definition of the critical density.

If, as in Sect.  \ref{empty}, we consider a vanishing density $\Omega^*_m  \rightarrow 0$ for $k=-1$,  
 this last equation tends to  $ 0 + \Omega_k + \frac{2}{1+\sqrt{2}}(1-0)= 1$. It
implies $\Omega_k = 3-2 \, \sqrt{2} =0.1716 $, thus according to (\ref{defo}) we get $H_0 = (1+\sqrt{2})/t_0$ 
in agreement with  the previous derivation (\ref{Hovide}).
  This is also consistent in the case $k=0$, introducing
(\ref{hkmin}) in (\ref{Omega}) implies for $\Omega_{\mathrm{k}}=0$
for a zero density.\\

It will certainly be very  useful for future comparisons with observational values to also  consider the usual definition of the critical density,
defined as  in the framework of Friedman's models (this density is indicated without a *),

\begin{equation}
\Omega_{\mathrm{m}}= \frac{\varrho}{\varrho_{\mathrm{c }}}\, \quad  
\mathrm{with} \quad  \varrho_{\mathrm{c }}= 3 \, H^2/(8 \pi G) \, .
\label{oprime}
\end{equation}

\noindent
  From the definition (\ref{roc}),  the two density parameters are related by

\begin{equation}
\Omega_{\mathrm{m}} = \Omega^*_{\mathrm{m}} \left( 1 - \frac{2}{H t} \right) \, .
\label{relom}
\end {equation}

\noindent
The relation between these two $\Omega$-parameters will be studied from numerical models.
With $\Omega_{\mathrm{m}}$,   relation (\ref{Omega}) becomes simply, 

\begin {equation}
\Omega_{\mathrm{m}} \, + \, \Omega_{\mathrm{k}} \, +  \Omega_{\lambda} = \, 1  \, ,
\label{Omegapr}
\end{equation}
\begin{equation}
\mathrm{with} \quad  \quad\Omega_{\lambda}  =  \frac{2}{ H \, t} \, .
\label{ol}
\end{equation}
\noindent
These two relations can also be derived directly by dividing (\ref{E1}) by $H^2$ and using (\ref{lambdot}). It also
corresponds to the above relation (\ref{smm}). 
There, $\Omega_{\mathrm{m}}$ is defined by (\ref{oprime}), $\Omega_{\mathrm{k}}$ by (\ref{defo}) and $\Omega_{\lambda}$
by (\ref{ol}).
The  above relations evidently also apply at time $t_0$, with the appropriate $H_0$ and  $t_0$.
  As  mentioned above,
$\Omega_{\lambda}$  must necessarily be  smaller than  $1$  for models with $k= 0$ and than 0.828 for  models with $k=-1$. 
  It is a fortunate circumstance that  an  equation  of the  form of (\ref{Omegapr}) is also valid
in scale invariant cosmology. The  difference  with the standard case is that the term 
 $\Omega_{\lambda}$ arising from scale invariance has replaced the usual term $\Omega_\Lambda$ due to
 the cosmological constant or dark energy.  This term arises naturally from scale invariance and does not demand the existence of  unknown particles.

We may also write    relation  (\ref{relom})
between the two density  parameters as follows,
\begin{equation}
\Omega_{\mathrm{m}} = \Omega^*_{\mathrm{m}} (1- \Omega_{\lambda}) \, .
\end{equation}
\noindent
An equivalent and useful  form is also
\begin{equation}
 \Omega^*_{\mathrm{m}}=  \frac{ \Omega_{\mathrm{m}}}{ \Omega_{\mathrm{m}}+ \Omega_{\mathrm{k}}} \, .
\label{omok}
\end{equation}

\noindent
These expressions allow us to make  some further remarks on the  $\Omega$-parameters:\\

 {{Case $k = 0$:}} 
\vspace{-1.5mm}
\begin{enumerate}
\item From (\ref{omok}), since $\Omega_{\mathrm{k}}=0$, we have $\Omega^*_{\mathrm{m}}=1 $ and this applies at all times in a model.
\item The ratio  $2/(t \, H)$ is  not a constant (except for empty models, Sect. \ref{empty}) and thus $\Omega_{\mathrm{m}}$ is not a constant according to (\ref{relom}),  it varies with   age in a given model, see also Sect. (\ref{w3}) for the detailed behavior  the $\Omega$-parameters in the past.
The balance between  $\Omega_{\lambda}$ and $\Omega_{\mathrm{m}}$ changes with time.
\item  In Friedman's models, there is only one model corresponding to $k=0$: the model with the critical density.
In the scale invariant framework, for  $k=0$ the fact that   $\Omega^*_{\mathrm{m}} = 1$
 does not imply specific values of  $\Omega_{\mathrm{m}}$ and  $\Omega_{\lambda}$.
Thus,  the additional term in (\ref{E1}) may  lead to
    a variety of possible models for $k=0$ with different  parameters $\Omega_{\mathrm{m}}$  and $\Omega_{\lambda}$
at time $t_0$. 
\item  According to its definition  (\ref{ol})  $\Omega_{\lambda}$ is positive, thus in order to satisfy (\ref{Omegapr}) for $k=0$,
 $\Omega_{\mathrm{m}}$ must be smaller than 1. Thus, the variety of models  for $k=0$ consists in models with
$\Omega_{\mathrm{m}}< 1$.\\ 

{{Case $k = \pm 1$:}}  
\vspace{1mm}
\item From (\ref{omok}),  $\Omega^*_{\mathrm{m}}$ is necessarily different from 1. 
\item According to their definitions,
the terms   $\Omega_{\mathrm{m}}$, $\Omega_{\lambda}$ and $\Omega_{\mathrm{k}}$ are expressed as functions 
of quantities, like $R(t)$, $H$, $t$,   that  change over the ages, thus these $\Omega$-terms are not constant in time,
(their behavior is examined in Sect. (\ref{w3}) on the basis of the conservation laws).
\item For $k=-1$, $\Omega_{\mathrm{k}}$ is positive as well as   $\Omega_{\lambda}$, thus the variety of possible models 
must have   $\Omega_{\mathrm{m}} < 1$.
\item For $k=1$,  if  $(\Omega_{\lambda}  +\Omega_{\mathrm{k}}) >0$, there is only a  variety of $\Omega_{\mathrm{m}}$-values 
smaller than 1. At this stage, we do not know the predicted range for  the various $\Omega$-parameters, but numerical results
will confirm that the sum $(\Omega_{\lambda}  +\Omega_{\mathrm{k}})$ is always positive for models with $k=1$..
\end{enumerate}

Depending on the values 
of $\Omega_{\mathrm{m}}$ and $\Omega_{\mathrm{k}}$, the displacements associated to scale invariance could provide
an important contribution $\Omega_{\lambda}$ to the energy-density present in the Universe. The CMB observations 
 \citep{deBern00}, WMAP \citep{Benn03} and the  \citet{Planck15} support  a flat model Universe with $k \approx 0$  with  
 $\Omega_{\mathrm{m}} \approx 0.30$ and  $\Omega_{\Lambda} \approx 0.70$. We note that  this last value  is   below the 
above permitted limit for $\Omega_{\lambda}$.  These values together with 
equation $\Omega_{\mathrm{m}}+\Omega_{\mathrm{k}}+\Omega_{\lambda} =1 $ would indicate that the energy associated to 
the effects resulting from scale invariance make a sizable fraction of the energy density of the Universe.


\subsection{The geometry parameters}  \label{geom}

We now consider  the geometry parameters $k$, $q_0 = - \frac{\ddot{R}_0 \, R_0}{\dot{R}^2_0}$
 and their relations with $\Omega_{\mathrm{m}}$, $\Omega_{\mathrm{k}}$ and $\Omega_{\lambda}$ at the present time $t_0$.
Expression (\ref{E2}) gives at the present time $t_0$, if the  pressure is zero,

\begin{equation}
\frac{k}{R^2_0} -2 \, q_0 H^2_0 +H^2_0 - 4 \, \frac{H_0}{t_0} = 0 \, .
\label{g1}
\end{equation}
 
\noindent
Divided by $H^2_0$ and with (\ref{defo}), this becomes

\begin{equation}
-2 \, q_0 +1 - \Omega_{\mathrm{k}}= \frac{4}{H_0 t_0} \, .
\label{qk}
\end{equation}

\noindent
Eliminating $\Omega_{\mathrm{k}}$ between (\ref{qk}) and (\ref{Omega}), we   obtain

\begin{equation}
2 \, q_0 =  \Omega^*_{\mathrm{m}}- \frac{2}{H_0 t_0} ( \Omega^*_{\mathrm{m}}+1)  \, ,
\label{int1}
\end{equation}

\noindent
and  thus, if we use $\Omega_{\mathrm{m}}$ rather  than $\Omega^*_{\mathrm{m}}$,

\begin{equation} 
q_0 \,= \, \frac{\Omega_{\mathrm{m}}}{2}-  \frac{\Omega_{\mathrm{\lambda}}}{2} \, \, .
\label{qzerom}
\end{equation}

\noindent
This  establishes relations between the acceleration parameter $q_0$ and the  expressions of the matter content for a 
scale invariant cosmology. If $k=0$ and thus  $ \Omega^*_{\mathrm{m}}=1 $, we have  from ({\ref{int1})

\begin{equation}
q_{0} = \frac{1}{2} -\Omega_{\lambda} \, =  \, \Omega_{\mathrm{m}} - \frac{1}{2} \, ,
\label{qkk}
\end{equation}
which provides a very simple relation between basic parameters. For a present $\Omega_{\mathrm{m}}=0.30$, we get 
$q_0=-0.20$.
Such relations could also be considered at  epochs different from the present one.
 
Most interestingly, the above basic relations are different from those of the $\Lambda$CDM. This is evidently expected 
since the basic equations (\ref{E1}) - (\ref{E3}) are different. Let us recall
that in the $\Lambda$CDM model with $k=0$ one has

\begin{equation}
q_{0}\, = \, \frac{1}{2}\Omega_{\mathrm{m}} -\Omega_{\Lambda} \, =  \, \frac{3}{2}\Omega_{\mathrm{m}} - 1 \, = \,
\frac{1}{2} -\frac{3}{2}\Omega_{\Lambda}  \, , 
\label{qlambda}
\end{equation}

\noindent
which may also be applied at different epochs. For $\Omega_{\mathrm{m}}=0.30$, we get 
$q_0=-0.55$.
In both cosmological, one has very simple relations expressing 
the $q$ parameter. However, these expressions lead to  significantly different results.\\

Let us now turn to the curvature parameter $k$. From the basic equation (\ref{E1}), we get

\begin{equation} 
\frac{k}{R^2_0} =H^2 _0\left( \frac{8 \, \pi G \varrho_0}{3 \, H^2_0} -1 + \frac{2}{t_0 H_0} \right) \, .
\label{basicpo}
\end{equation}

\noindent
With the definition of the critical density (\ref{roc}) and with ({\ref{ol}), this becomes at the present time $t_0$,

\begin{equation}
\frac{k}{R^2_0} =H^2 _0\left[ (\Omega^*_{\mathrm{m}} -1) \left(1 - \frac{2}{t_0 H_0} \right )\right]  \, , 
\label{kom}
\end{equation}

\noindent
 which establishes a relation between $k$ and $\Omega*_{\mathrm{m}}$.  It confirms that
 if $\Omega^*_{\mathrm{m}}=1 $, one also has $k=0$
and reciprocally. We also verify that for  $2/(t_0 \, H_0)=1$, we effectively have $k=0$ in agreement with (\ref{hkmin}). Values of $\Omega^*_{\mathrm{m}} > 1$ give a positive $k$-value,
 values smaller than $1$ give a negative $k$-value.  Using  $\Omega_{\mathrm{m}}$,
the above relation also writes at present time,

\begin{equation}
\frac{k}{R^2_0} =H^2 _0\left[ \Omega_{\mathrm{m}}  -\left(1- \frac{2}{t_0 H_0} \right )\right]  \, , 
\label{kom2}
\end{equation}

\noindent
which is finally just equivalent to (\ref{Omegapr}) at time $t_0$. 

We also have
a relation between $k$ and $q_0$. From (\ref{int1}), we get
 
\begin{equation}
\Omega^*_{\mathrm{m}}=\frac{2q_0+\frac{2}{H_0t_0}}{1-\frac{2}{H_0 t_0}}  \, ,
\end{equation}
\noindent
and using this in (\ref{kom}), we obtain 

\begin{equation}
\frac{k}{R^2_0} = H^2_ 0 \left[ 2 \, q_0 -1 + \frac{4}{H_0 t_0} \right] \, .
\label{kq}
\end{equation}

\noindent 
For $k=0$, it evidently gives the same relation as from relation (\ref{qzerom}) above. We again emphasize that in all these expressions
$t_0$ is not the present age of the Universe, but just  the present time  in a scale where $t_0 = 1$. As in Sect. \ref{empty}, 
the present age $\tau= t_0 -t_{\mathrm{in}}$, where the values of the initial time $t_{\mathrm{in}}$ depend on the considered model.


\subsection{Inflexion point in the expansion} \label{inflex}

The Friedman models do not have an inflexion point, the second derivative $\ddot{R}$ is always negative and thus $q$ is positive at all times. 
In the scale invariant cosmology, like in the  $\Lambda$CDM models, there are both a braking force of gravitational attraction and 
an acceleration force acting in the Universe model. There may thus be epochs  dominated by gravitational braking and other epochs  by acceleration.
According to (\ref{E3}), there is an inflexion point in the curves $R(t)$, when we have the equality of  braking and acceleration,

\begin{equation}
\frac{4 \pi G}{3}(3p+\varrho) = \frac{H}{t} \,  .
\label{infl}
\end{equation} 

\noindent
We better use  (\ref{qzerom}), valid at any epoch $t$. For $q=0$ in the scale invariant models,   an inflexion in the curve $R(t)$ 
occurs at time $t$ when

\begin{equation}
 \Omega_{\mathrm{m}} \, =  \, \Omega_{\lambda}  \, .
\label{oml}
\end{equation}

\noindent
An inflexion point occurs when there is an equilibrium between these two $\Omega$-parameters.  The gravitational term dominates
in the early epochs and the $\lambda$-acceleration dominates in more advanced stages. The higher the $\Omega_{\mathrm{m}}$-value,
the later the inflexion point occurs. The empty model
discussed  in Sect. \ref{empty}, where $R(t) \sim t^2$, seems to be an exception. It shows  no inflexion point in the course of evolution and is  starting with an horizontal tangent, before
 accelerating continuously. A non-zero density may lead to  positive 
values of $q$ at the origin followed by  negative ones after the inflexion point. At this stage, we do not know whether 
scale invariant models predict an explosive origin. 

For a flat model with $k=0$, we can further precise the location of the inflexion point. Since in this case, $ \Omega_{\mathrm{m}} =
1- \Omega_{\lambda} $, we have at the inflexion point

\begin{equation}
(1- \Omega_{\lambda}) \, = \,  \Omega_{\lambda}\, , \quad  \quad \mathrm{and \; thus} \quad \Omega_{\lambda}
= \Omega_{\mathrm{m}} = \frac{1}{2}  \, .
\label{ql}
\end{equation}

\noindent
The matter and the $\lambda$-contributions should be equal and both equivalent to 1/2.
With (\ref{ol}), the inflexion point for $k=0$ models occurs at time $t$ such that
\begin{equation}
t \, = \, \frac{4}{  H } \, ,
\label{t4}
\end{equation}
\noindent
where $t$ is counted in the scale where $t_0=1$ at present and the same for $H$.

These results differ from those for the $\Lambda$CDM models.  According to (\ref{qlambda}), we
have $q=0$   for a flat $\Lambda$CDM model  when \citep{Suther15},

\begin{equation}
 \frac{1}{2}  \,  \Omega_{\mathrm{m}} \, =  \, \Omega_{\Lambda}  \, .
\label{llml}
\end{equation}

\noindent
This is to be compared to  the scale invariant case  given  by (\ref{oml}). The acceleration term  needs only to reach one half
of the gravitational term to reach the critical limit   in the  $\Lambda$CDM model, while   in the scale invariant case the inflexion
point is reached for the equality of the two terms. This may provide  possible observational tests, since the existence 
of an inflexion point in the evolution of the expansion factor $R(t)$ has been analyzed in several recent works \citep{Melch07,Ishida08,Suther15,Viten15,Moresco16}.

\section{Conservation laws}

\subsection{General expression}

The laws of conservation are  fundamental properties of physics.
 It is clear that including a new invariance such as the  scale invariance  will influence in some
way the  laws of conservation.  In addition, we have also explicitly accounted
for the scale invariance of the vacuum at the macroscopic scales by using the differential equations (\ref{diff1}) and  (\ref{diff2}). These various hypotheses have an impact on the conservation laws.
We derive the conservation laws from  the basic  equations (\ref{E1}) to (\ref{E3}).
We first rewrite  (\ref{E1}) as follows and take its derivative,

\begin{eqnarray}
8 \, \pi G \varrho R^3 = 3\, k R+ 3 \, \dot{R}^2 R+ 6  \, \frac{\dot{\lambda}} {\lambda} \dot{R}  R^2 \, . \\  \vspace{3mm}
\frac{d}{dt} (8 \, \pi G \varrho R^3 ) 
 = 3\, k\dot{ R}+ 3 \, \dot{R}^3 +6 \, \dot{R}\ddot{R}R+  \\ \nonumber
 + 6 \,  \ddot{R} R^2\frac{\dot{\lambda}}{\lambda}+ 6 \,  \dot{R} R^2\frac{\ddot{\lambda}}{\lambda}+ 
12 \,  \dot{R}^2  R \frac{\dot{\lambda}}{\lambda}-6 \,  \dot{R} R^2\frac{\dot{\lambda^2}}{\lambda^2} \,  \\ \nonumber
= -3 \dot{R} R^2 \left[-\frac{k}{R^2}- \frac{\dot{R}^2}{R^2}-2 \frac{\ddot{R}}{R}-  
2  \frac{\ddot{R}}{ \dot{R}} \frac{\dot{\lambda}}{\lambda}-2  \frac{\ddot{\lambda}}{\lambda}
- 4 \frac{\dot{R} \dot{\lambda}}{R \lambda} +2 \frac{\dot{\lambda}^2}{\lambda^2}
\right]  \, .
\end{eqnarray}

\noindent
We recognize terms belonging to the second member of (\ref{E2}), so that the above relation becomes

\begin{equation}
\frac{d}{dt} (8 \, \pi G \varrho R^3 )=-3\, \dot{R} R^2 \left[8 \, \pi G p-  2 \, \frac{\ddot{R}}{ \dot{R}} \frac{\dot{\lambda}}{\lambda}
-2 \, \frac{\ddot{\lambda}}{\lambda} +2 \frac{\dot{\lambda}^2}{\lambda^2}\right]  \, .
\end{equation}

\noindent
The scale invariance of the empty space imposes relations (\ref{diff2}), which leads to further simplifications

\begin{equation}
\frac{d}{dt} (8 \, \pi G \varrho R^3 )=-3\, \dot{R} R^2 \left[8 \, \pi G p-  2 \, \frac{\ddot{R}}{ \dot{R}} \frac{\dot{\lambda}}{\lambda}
-2 \frac{\dot{\lambda}^2}{\lambda^2}\right]  \, .
\end{equation}

\noindent
Using (\ref{E3}) again, the third of our fundamental equations,  to express the last two terms on the right of the above equation, we  obtain

\begin{equation}
\frac{d}{dt} (8 \, \pi G \varrho R^3 )=-3\, \dot{R} R^2 \left[8 \, \pi G p+\frac{{R}}{ \dot{R}} \frac{\dot{\lambda}}{\lambda}
\left(8 \, \pi Gp+ \frac{8 \, \pi G \varrho}{3}\right) \right]  \, .
\end{equation}

\noindent
We simplify by  $8 \, \pi G$, which is a constant, and write the above equation in differential form. With further simplifications it becomes,

\begin{equation}
3 \, \lambda \varrho dR + \lambda R d\varrho+ R \varrho d\lambda +3 \, p \lambda dR +3 p R d\lambda = 0 \, .
\end {equation}

\begin{equation}
 \mathrm{and} \; \;  \; 3 \, \frac{dR}{R} + \frac{d \varrho}{\varrho}+\frac{d \lambda}{\lambda}+ 3 \, \frac{p}{\varrho}\; \frac{dR}{R}+
3 \, \frac{p}{\varrho} \; \frac{d\lambda}{\lambda} = 0 \, .
\label{conserv1}
\end{equation}

\noindent
This can also be written in a form rather similar to the usual conservation law,

\begin{equation}
\frac{d(\varrho R^3)}{dR} + 3 \, p R^2+ (\varrho+3\, p) \frac{R^3}{\lambda} \frac{d \lambda}{dR} = 0 \, .
\label{conserv2}
\end{equation}

\noindent
These last two expressions are convenient forms of the law of  conservation of mass-energy  in the scale invariant cosmology. For a constant $\lambda$, we evidently recognize the
  conservation law or first integrals of the cosmological equations derived from General Relativity with the Robertson-Walker metric.


\subsection{Specific cases: matter, radiation and vacuum}  \label{w3}

We now apply the above equation of conservation to some specific media characterized by different equations of state.
We write the equation of state in the general form,

\begin{equation}
P \, = \, w \,  \varrho \, ,  \quad  ( \mathrm{with \;} c^2 =1) \,  ,
\label{etat}
\end{equation}

\noindent
where $w$ is  taken here as a constant, (variable $w$ depending on the epochs have been considered by some authors).
The equation of conservation (\ref{conserv1}) becomes

\begin{equation}
 3 \, \frac{dR}{R} + \frac{d \varrho}{\varrho}+\frac{d \lambda}{\lambda}+ 3 \, w \,  \frac{dR}{R}+
3 \, w \, \frac{d\lambda}{\lambda} = 0 \, ,
\label{conserw}
\end {equation}

\noindent
with the following simple integral which covers all possible cases of constant $w$,

\begin{equation}
\varrho \, R^{3(w+1)}  \,  \lambda ^{(3w+1)} \,= const. 
\label{3w}
\end{equation}

\noindent
Different $w$-values correspond to different types of medium. For $w=0$, we have the case of ordinary matter of density
$\varrho_{\mathrm{m}}$, exerting no pressure. 
We get 

\begin{equation}
\varrho_{\mathrm{m}} \, \lambda \, R^3 =const. 
\label{consm}
\end{equation}

\noindent
which means that the inertial  and gravitational mass within a covolume should both (in agreement with the Equivalence Principle)    slowly  
increase over the ages. At this stage, one may wonder how large are the changes of $\lambda$ over the life of the Universe. For the
empty model (Sect. 3), the change  of $\lambda$ is enormous, going from 1 at present to infinity at the origin.  In a more realistic model,
for example in a flat model with $\Omega_{\mathrm{m}}=0.30$, $\lambda$ varies from 1 at present to about 1.4938 at 
the origin situated at $0.66943 \; t_0$ (cf. Paper III).

Although  the effect of the variations of $\lambda$ appears very  limited in (\ref{consm}), how could we understand it? We do not expect any matter creation  as in Dirac's Large Number Hypothesis \citep{Dirac73}
 and thus  the number of baryons should be a  constant. However, 
as mentioned above, since an additional fundamental invariance has been
accounted for,  some changes in the conservation laws are necessarily to be expected.
We note that a change of the inertial and gravitational mass is not a new fact, it is well known in Special Relativity, where 
the masses change as a function of their velocity.   In the standard model of particle physics, the constant masses of elementary particles originate from the interaction of the Higgs field \citep{Higgs14,Englert14} in the 
vacuum with originally massless particles.
Here, the assumption of scale invariance of  the vacuum  (at large scales) 
and  of  the gravitation field would not let the mass invariant and make them to slowly slip over the ages, however by a limited amount
in realistic models.

We do not know whether the present scale invariant models correspond to Nature. 
Some  initial fundamental assumptions consistently lead to some consequences, however     comparisons between models
and observations  may possibly confirm or infirm these results. This is why in a further paper we will  proceed to model constructions and make such comparisons.

For now, we may check 
 that the above expression (\ref{consm}) is   fully  consistent with the hypotheses made.
 From relation (17) derived from the study of the momentum-energy tensor in Paper I,  we  obtained   that $\varrho' \lambda^2 = \varrho$, where we recall that the prime refers to the value in  General Relativity and the symbols without a prime apply to the values in the scale invariant system.
Thus expression (\ref{consm}) becomes,  also accounting for the scale transformation  $ \lambda \, R=R'$, 

\begin{equation} 
\varrho_{\mathrm{m}} \, \lambda \, R^3  =  \varrho' \, \lambda^3 \, R^3 =  \varrho' \, R'^3=const.
\label{massconsE}
\end{equation}
\noindent
This is just the usual mass conservation law  in General Relativity. 

Let us go on with the conservation law for relativistic particles and in particular for radiation with density $\varrho_{\gamma}$. 
Here,  the ratio $w$  of pressure to energy density  is $w=1/3$.  From the equation of conservation (\ref{3w}),
we get

\begin{equation}
\varrho_{\gamma} \, \lambda^2 \, R^4 =const.   
\label{consrad}
\end{equation}
\noindent
There, a term $\lambda^2$ intervenes.
As for the mass conservation, we may check its consistency with General Relativity. Expression (\ref{consrad}) becomes 
$\varrho'_{\gamma} \, \lambda^4 \, R^4 =const. $ and
thus $\varrho'_{\gamma} \, \, R'^4 =const. $ in  the Einstein framework.

Another interesting  case is that of the vacuum or
dark energy (if any one) with density $\varrho_{\mathrm{v}}$.  It would obey to the equation 
of state $ p= -\varrho$ with $c=1$. Thus, we have $w=-1$ and
 (\ref{3w}) becomes

\begin{equation}
\varrho_{\mathrm{v}} \lambda^{-2} = const.
\label{consv}
\end{equation}
\noindent
suggesting a decrease of the vacuum energy over the ages.
With $\varrho'_{\mathrm{v}} \lambda^2 = \varrho_{\mathrm{v}}$, this   corresponds to  $\varrho'_{\mathrm{v}}= const.$ in  the Einstein framework.    This  is the standard result, which corresponds to the presence of a cosmological constant in General Relativity.\\

We may now examine the time evolution of the $\Omega$-parameters in a given model, in complement of the remarks in Sect. \ref{dens}.
In the matter dominated era, we have  since $\varrho \sim t/R^3$ and $\varrho_{\mathrm{c}} \sim H^2$,

\begin{equation}
\Omega_{\mathrm{m}} \, \sim \, \frac{t}{R^3  H^2} \, .
\end{equation}
\noindent
For $\Omega_{\mathrm{k}}$ and $\Omega_{\lambda}$, the behaviors are like

\begin{equation}
\Omega_{\lambda} \, \sim \, \frac{1}{tH} \quad \mathrm{and}  \quad \Omega_{\mathrm{k}} \, \sim \, \frac{1}{R^2 \, H^2} \, .
\end{equation}
\noindent
We remark that these three $\Omega$-parameters would stay constant in time, 
only if the expansion factor $R(t)$ would go like
$t$. This is   the case neither  for the empty model, nor for the models 
with some matter-density  since they have both a braking and an acceleration phase.
 This confirms that the three $\Omega$-parameters vary in time in  scale invariant models,
(evidently for $k=0$ one has $\Omega_{\mathrm{k}}=0$ at all times). 
 Let us now turn to
the parameter  $\Omega^*_{\mathrm{m}}$. We have seen that it is equal to 1 and remains constant in models with $k=0$.
What about the models with $k = \pm 1$?  Let us examine the scaling  predicted from  (\ref{omok}),

\begin{equation}
 \Omega^*_{\mathrm{m}}=  \frac{ 1}{ 1+ \frac{\Omega_{\mathrm{k}}}{\Omega_{\mathrm{m}}}} \, \sim \,
\frac{1}{1+ \frac{R}{t}} \, .
\end{equation}
\noindent
There also $R$ should go like $t$ to maintain the constancy in time of $\Omega^*_{\mathrm{m}}$. As this is not the case,
we conclude that $\Omega^*_{\mathrm{m}}$ also varies with time in the models with $k= \pm 1$.\\

The above conservation laws   are necessary for   establishing the past matter and radiation history of the Universe, as well as for
 the integration of the cosmological equation (\ref{E1}).  They will allow us 
to consider some terms   as constant during the integration of the equations over the ages.

\section{Conclusions}

We have derived  the equations of cosmologies in the scale invariant framework, also accounting  for the scale invariance of the 
vacuum at large scales. This hypothesis brings interesting simplifications in the equations. On the whole, the scale invariant equations of cosmology  only contain  one 
additional term compared to the standard equations derived from General Relativity. The main physical consequence of this additional term is an acceleration of the cosmic expansion.

We first considered the model  of a zero-density Universe. While in Friedman's models, the expansion of such a Universe model 
behaves like  $R(t) \sim  t $, in the scale invariant framework the model shows an accelerated expansion going  like $t^2 $.

 The main conclusion is that the contribution $\Omega_{\lambda}$ due to the effects of  scale invariance to the energy density of the
Universe is an important one, with  $\Omega_{\mathrm{m}}+\Omega_{\mathrm{k}}+\Omega_{\lambda} =1 $.
This energy density is in the form of the accelerated expansion.
If this happens to apply, we might wonder about the  need to invoke unknown particles.

For  zero curvature $k=0$, there is a whole family of models with different possible density parameters
 $\Omega_{\mathrm{m}} < 1$.
The geometrical parameters of the models  and their relations with the matter-density are also studied. 
The non-empty scale invariant models have an inflexion point  with $q=0$  in their evolution $R(t)$: 
 there is first  a gravitational braking of
the expansion followed by a cosmic acceleration. The conditions for the inflexion point are not the 
same as for the $\Lambda$CDM models. The inclusion of the scale invariance
 modifies the conservation laws, which thus show a factor depending on the cosmic time.

On the whole, a consistent framework appears to exist for scale invariant cosmology.
The observational tests will tell us whether this framework is worth to be further explored.

\vspace*{5mm}

\noindent
Acknowledgments: I want to express my best thanks to the physicist D. Gachet and Prof. G. Meynet for their 
continuous encouragements.

\appendix

\section{Appendix: derivation of the basic equations}

We derive the scale invariant equations in a straightforward way. Instead of applying  the Robertson-Walker metric to
 the scale invariant field equation, we directly apply the scale transformations to the  differential equations of cosmologies
in the system of General Relativity. These equations are
   
\begin{equation}
\frac{8 \, \pi G \varrho' }{3} = \frac{k}{R'^{2}}+\frac{ \dot{R}'^{ 2}}{R'^{ 2}}-\frac{\Lambda_{\mathrm{E}}}{3}  \, \; ,
\label{E111}
\end{equation} 

\begin{equation}
-8 \, \pi G p'  = \frac{k}{R'^{2}}+ 2 \frac{\ddot{R'}}{R'}+ \frac{\dot{R}'^{ 2}}{R'^{2}}- \Lambda_{\mathrm{E}} \, .
\label{E222}
\end{equation}
\noindent
There, $\Lambda_{\mathrm{E}} $ is the Einstein cosmological constant, $G$ is the gravitational constant which is a real constant,
 $k$ is the curvature parameter which may take values $0$ and $\pm 1$,  $p'$ and $\varrho'$ are the 
pressure and density in the system of General Relativity coordinates.
Now,  we make the transformations

\begin{equation}
 R'\,  = \, \lambda \, R \quad  \quad \mathrm{and} \quad dt' \,  = \, \lambda \, dt \, . 
\end{equation}
\noindent
We  get

\begin{equation}
\dot{R}' \, = \, \frac{dR'}{dt'}= \frac{\dot{\lambda} \, R+ \lambda \, \dot {R}}{\lambda} \, ,
\end{equation}
\noindent
where the dot over a symbol indicates its derivative with respect to the time "t" in the scale invariant system. Then, we have

\begin{equation}
\frac{\dot{R'}}{R'} \,= \,  \frac{1}{\lambda} \, \left(\frac{\dot{\lambda}}{\lambda}+ \frac{\dot{R}}{R} \right) \, .
\label{rr}
\end{equation}
\noindent
The second derivative  $\ddot{R'}$ becomes

\begin {equation}
\ddot{R'} = \frac{d (\frac{dR'}{dt'})}{dt} = \frac{1}{\lambda^2}  (\ddot{\lambda}R+ 2 \dot{\lambda} \dot{R} +
\lambda \ddot {R}) -  \frac{(\dot{\lambda} \, R+ \lambda \, \dot {R})}{\lambda^2}  \frac{\dot{\lambda}}{\lambda} \, ,
\end{equation} 
\noindent
and
\begin{equation}
\frac{\ddot{R'} }{R'} \, = \,  \frac{1}{\lambda^2} \, \left(\frac{\ddot{\lambda}}{\lambda}+
 \frac{\dot{\lambda} \, \dot{R}}{\lambda \, R}+ \frac{\ddot{R}}{R} - \frac{\dot{\lambda}^2}{\lambda^2} \right) \, .
\end{equation}

\noindent
Thus, by replacing in (\ref{E111})    we obtain

\begin{equation}
\frac{8 \, \pi G \varrho }{3} = \frac{k}{R^2}+
\frac{\dot{R}^2}{R^2}+ 2 \, \frac{\dot{\lambda} \, \dot{R}}{\lambda \, R}+\frac{\dot{\lambda}^2}{\lambda^2} - \frac {\Lambda_{\mathrm{E}} \lambda^2}{3} \,
\label{Ae1}
\end{equation}
\noindent
and  from (\ref{E222})  after simplifications,

\begin{equation}
-8 \, \pi G p = \frac{k}{R^2}+ 2 \frac{\ddot{R}}{R} + 2 \frac{\ddot{\lambda}}{\lambda}+\frac{\dot{R}}{R}^2
+ 4 \frac{\dot{R} \, \dot{\lambda}}{R \, \lambda}-\frac{\dot{\lambda^2}}{\lambda^2} -\Lambda_{\mathrm{E}} \,  \lambda^2  \, .
\label{Ae2}
\end{equation}

\noindent
The various quantities in the equations are expressed in the general system where  scale invariance is a property.
There, we have used  the relations (17) of Paper I imposed by the scale invariance of the energy-momentum tensor,
$p= p' \,  \lambda^2$ and $\varrho = \varrho' \, \lambda^2$. These two relations correspond to the results by 
\citet{Canu77}. At this stage, these relations do not account for the relations expressing the scale invariance of the empty space,
which lead to substantial simplifications.

\bibliographystyle{aa}
\bibliography{Maeder II-AA}
\end{document}